\documentclass[pre,a4paper,superscriptaddress,twocolumn,showpacs,amsmath,amssymb,floatfix]{revtex4}

\usepackage{graphicx}
\usepackage{amssymb}
\usepackage{float}
\usepackage{hyperref}

\usepackage{color}

\begin{document}
	
\title{Dynamical thermalization and turbulence in social stratification models}
	
\author{Klaus M. Frahm}
\affiliation{{\mbox Univ Toulouse, CNRS, Laboratoire de Physique Th\'eorique, 
Toulouse, France}}
\author{Dima L. Shepelyansky}
\affiliation{{\mbox Univ Toulouse, CNRS, Laboratoire de Physique Th\'eorique, 
Toulouse, France}}
\affiliation{\mbox{{\bf Author to whom correspondence should be addressed: dima@irsamc.ups-tlse.fr}}}

\date{March 25, 2026}

\begin{abstract}
We study the nonlinear chaotic dynamics in a system of linear
oscillators coupled by social network links
with an additional stratification of
oscillator energies, or frequencies,
and supplementary nonlinear interactions.
It is argued that this system can be viewed as
a model of social stratification in a society
with nonlinear interacting agents
with energies playing a role of wealth states
of society. The Hamiltonian evolution is characterized 
by two integrals of motion being
energy and probability norm.
Above a certain chaos border the 
chaotic dynamics leads to dynamical thermalization
with the Rayleigh-Jeans (RJ) distribution
over states with given energy or wealth.
At low energies, this distribution
has RJ condensation of norm 
at low energy modes. We point out a similarity
of this condensation with the wealth
inequality in the world countries
where about a half of population owns
only a couple of percent of the total wealth.
In the presence of energy pumping and absorption, the system reveals
features of the Kolmogorov-Zakharov turbulence 
of nonlinear waves. 
\end{abstract}

%

\maketitle
	
{\bf Social stratification in a society is actively discussed in
  social sciencies \cite{marx,lenski,sanders,kerbo,wikistratif}
  and here we study a mathematical model of this phenomenon.
The wealth levels  of society  are associated with energy levels
of nearly constant density of states coupled by social network links between agents.
It is shown that  above a chaos border the nonlinear interactions between agents
lead to chaos induced dynamical thermalization with the Rayleigh-Jeans (RJ) distribution
over states. The distribution is characterized by temperature
and chemical potential related to two integrals of nonlinear dynamics
being system energy and the total probability norm. At low energy and temperature
there is a phenomenon of RJ condensation actively studied experimentally and
theoretically for light propagation in multimode optical fibers.
Due to condensation a significant norm fraction is concentrated
at lowest energy or wealth states. It is argued that
such a RJ condensation describes the existence of a huge phase
of poor households in a society and a tiny oligarchic phase
that owns a big fraction of total wealth. In presence of
energy injection at low energy states and absorption at high energies, 
a flow of energy or wealth from low to high energy modes appears which is 
similar to the  Kolmogorov-Zakharov turbulence spectra of nonlinear waves.
The Lorenz curves for wealth of households,
obtained from the RJ distributions, are shown to be
similar to those in the world countries. 
}

\section{Introduction} 
\label{sec1}

Social stratification is a society structure based on 
ranking of its members and groups
being classified on the grounds of their wealth, activity, education and
other related factors. This field is under active investigations in social
and economy sciences starting from the theory of society class structure  
of Karl Marx
(see e.g. \cite{marx,lenski,sanders,kerbo,wikistratif}).
Due to  its complexity it can be useful to propose a mathematical model 
of social stratification of society (SSS) and study its properties with 
tools of dynamical nonlinear systems.
One of the arguments for a construction of such SSS models is 
a recent  enormous development of various social networks that gained a
significant importance for communications, opinion formation and
interactions between  society agents 
(see e.g. \cite{dorogovtsev10,newmanbook}).
The properties of numerous social networks have been studied by various groups
as reviewed in \cite{dorogovtsev10,newmanbook}. However, all these studied
are based on a linear matrix algebra of links between network nodes 
provided by their
adjacency matrix of real systems. At the same time the interactions between
society members are very complex and it is evident that they should include
nonlinear interactions. Another feature, usually absent in studied social 
networks, is that commonly their adjacency matrix is composed only by 
off-diagonal links with zero diagonal
component while it is natural to expect that social stratification
should be related to the presence of a certain diagonal component
attached to e.g. wealth inequality in a society.

In this work, we introduce a mathematical SSS model of interacting agents
which has typical social network structure of links between agents
but in addition also includes a nonlinear interaction of agents and 
social stratification described by a linear diagonal term
added to an adjacency matrix.  The proposed SSS model
is associated to the following Hamiltonian:
\begin{align}
\label{eqIntegrals}
\mathcal{H}=\sum_{n,n'} \psi_n^* H_{n,n'}\psi_{n'}+\frac{\beta}{2}
\sum_n |\psi_n|^4  \; . 
\end{align}
which describes the dynamical evolution of a 
nonlinear field $\psi_n$ for agents $1 \leq n \leq N$. 

In fact, this is 
a classical Hamiltonian of an oscillator system  
if we present complex amplitudes of agent $n$ as $\psi_n=(q_n+ip_n)/\sqrt{2}$
with canonical coordinates $q_n$ and $p_n$ of coordinate and momentum.
Then the time evolution of the system is given by
\begin{align}
\label{eqNLeq1}
i\frac{\partial\psi_n(t)}{\partial t}&=\sum_{n'=1}^N H_{n,n'} \psi_{n'}(t) 
+   \beta \vert\psi_n(t)\vert^2\psi_n(t) 
\end{align}
where the parameter $\beta$ describes the strength of the 
nonlinear perturbation and $H_{n,n'}$ are the matrix elements of a real 
symmetric matrix $H$ which will be specified more explicitly in the 
next Section. For $\beta=0$, (\ref{eqNLeq1}) represents a quantum 
Schr\"odinger equation or a simple coupled linear oscillator system 
where different oscillator modes are coupled by the matrix $H$.

The dynamical system (\ref{eqNLeq1}) has two integrals of motion which 
are the quantity $\eta =\sum_n |\psi_n|^2$ (or weighted number 
of agents), called {\em norm} \footnote{
For simplicity, we simply call the quantity $\eta$ 
``norm'' even though it actually corresponds to the ``squared norm''
$\|\psi\|^2=\langle\psi|\psi\rangle=\sum_n|\psi_n|^2$ if 
$|\psi\rangle=\sum_n \psi_n|n\rangle$ is viewed as a ``quantum state''
with time evolution (\ref{eqNLeq1}) if $\beta=0$. 
}, and the 
{\em classical oscillator  energy} $\mathcal{H}$ in (\ref{eqIntegrals}).
The norm $\eta$ is an integral of motion
since the Hamiltonian (\ref{eqIntegrals}) is Hermitian
and the time evolution from (\ref{eqNLeq1}) is unitary
thus preserving the norm $\eta$. 
Mathematically, it is a 
straightforward calculation from (\ref{eqNLeq1}) to verify that 
both $\eta(t)=$const. and $\mathcal{H}(t)$=const for any $\beta$.
Here, we fix the norm to $\eta=1$. 
The case $\eta\neq 1$ can be reduced to the 
case $\eta=1$ by a suitable rescaling of 
$\psi$ and $\beta$. 

In the SSS model, which will be defined more precisely 
in the next Section, the nonlinear fields $\psi_n$ describe
$N$ society agents, nondiagonal elements $H_{n,n'}$
represent social network links between agents
with nonlinear interactions induced by the $\beta$-term
while the diagonal elements $H_{n,n}$
characterize social stratification.

According to the general properties of
dynamical chaos \cite{arnold,sinai,chirikov1979,lichtenberg},
we expect that at a moderate value of the nonlinear parameter $\beta$ 
above a certain chaos border $\beta_{ch}$,
the dynamics becomes globally chaotic
with positive maximal Lyapunov exponent and
positive Kolmogorov-Sinai entropy.
Below the chaos border the dynamics is integrable according to the
Kolmogorov-Arnold-Moser (KAM) theorem 
\cite{arnold,sinai,chirikov1979,lichtenberg}.
The chaotic dynamics should lead to
dynamical thermalization of the system to a state of maximal 
entropy on the phase space defined by the two integrals of motion. 

This particular case leads to the Rayleigh-Jeans (RJ) thermal 
distribution over linear eigenmodes of the matrix $H$ 
\cite{landau,zakharovbook}:
\begin{equation}
\rho_m = \frac{T}{E_m-\mu} \; ({\rm RJ}) 
\label{eqrj}
\end{equation}
where $\rho_m$ is the average occupation number of the oscillator 
mode at energy $E_m$; $T$ and $\mu$ are the temperature and chemical 
potential determined by the values of the two integrals of motion 
(see next Section for more technical details).

We stress that in our model the RJ thermalization 
has a purely dynamical origin due to the chaotic dynamics 
and that there is no contact with a thermal bath. 
The issue of dynamical thermalization in such oscillator chains 
(with nonlinear perturbations) has a long and complicated history 
with the first numerical experiments 
performed by Fermi-Pasta-Ulam-Tsingu (FPUT)
on the MANIAC I computer in 1955  with the conclusion that
``The results show very little, if any,
tendency toward equipartition of energy between the degrees of freedom''
\cite{fpu1955}.
Later it was shown that the absence of thermalization in the FPUT model 
was due to small nonlinear values of couplings between oscillators
and the system proximity to completely integrable nonlinear models
e.g. like the Toda lattice. At a stronger nonlinearity,
above the chaos border,
the dynamical thermalization was recovered
(see the extensive literature about the FPUT problem e.g. in
\cite{chirikovfpu1,chirikovfpu2,livi,benettin,fpu50} and Refs. therein).
The FPUT model has only one integral of
energy so that in the thermalized regime the average energy of
each linear mode is equal to temperature $T$ ($\mu=0$ in  (\ref{eqrj})).

In \cite{rmtprl}, it was argued that the proximity to integrable models
can be excluded for the case when the matrix $H$ of linear couplings 
between harmonic oscillator modes, in Eqs.
(\ref{eqIntegrals}), (\ref{eqNLeq1}), 
is random according to Random Matrix Theory (RMT) 
which  describes the universal properties of spectra and eigenstates
of complex nuclei, atoms and molecules \cite{wigner,mehta}.
It was shown in \cite{rmtprl} that for an RMT matrix $H$ 
a moderate nonlinearity $\beta$
above a certain chaos border chaos indeed leads to RJ thermalization 
with the distribution (\ref{eqrj}) 
both for positive and negative temperatures $T$.

The RJ thermalization has also been established for a more realistic 
model of light propagation in a quantum chaos fiber,
described by a nonlinear Schr\"odinger equation (NSE),
if the nonlinearity is sufficiently strong \cite{ourfiber}. 
In such a fiber the NSE depicts the propagation along a fiber,
with the longitudinal coordinate $z$
corresponding to time, and the D-shape of the fiber cross-section leads to
quantum chaos of linear eigenmodes creating RMT type properties
of the matrix $H$ in (\ref{eqIntegrals}).

We point out that RJ thermalization has also been actively studied 
numerically and experimentally in the context of 
light propagation in multimode optical fibers 
\cite{picozziphrep,wabnitz,picozzi1,picozzi2,babin,chris,picozzi3}.
At the same time the origins of this thermalization were attributed to
the Kolmogorov-Zakharov (KZ) turbulence
\cite{zakharovbook,nazarenkobook,nazarenkobook,galtier}
without referring to chaos and KAM integrability. 
At low temperatures the RJ distribution may have an enormous norm 
concentration at the ground state mode and the emergence of 
this {\em RJ condensation} was established numerically in \cite{picozzi1}
and experimentally in \cite{wabnitz,picozzi2}.

We may also note that the Fr\"ohlich condensate
for biomolecules at room temperature,
proposed in \cite{frohlich1,frohlich2}, 
has certain similarities with the RJ condensate
showing analytically that the RJ distribution (\ref{eqrj})
has a huge norm concentration at the lowest frequency mode
even though in \cite{frohlich1,frohlich2} the considered system includes
external pumping and dissipation
(see also the discussion in \cite{ourfiber}).

The RJ thermalization observed in \cite{rmtprl} 
is due to the combination of the nonlinear term and the 
RMT structure of $H$ which 
is very different from the link structure of social networks
described by very sparse matrices with
small-world properties \cite{dorogovtsev10,newmanbook}. 
In \cite{wth}, dynamical RJ thermalization was discussed 
for such a case where the matrix $H$ 
is constructed from a real social network 
but due to the absence of diagonal terms (absence of self links) 
this model has no social stratification. 

In this work, we consider a more realistic SSS model 
with social network links and also with
a social stratification modeled by additional diagonal 
matrix elements in $H$ and where
nonlinear interactions and dynamical chaos
also lead to RJ thermalization and condensation 
for suitable choices of system parameters.
Here the linear eigenenergies $E_m$ have a rather constant 
density of states. 
For a society such energy levels $E_m$ (or oscillator frequencies)
can be considered as certain levels of
wealth of society agents
with low and high energies
corresponding to poor and rich people in a human society.
Then the RJ condensation at low energy states
can be considered as a phase of
poor population at low wealth
while a population fraction at high energies (or wealth)
can be viewed
as an oligarchic phase of very rich people.
We highlight a certain similarity between RJ condensation
in these SSS models 
and the wealth inequality in the human society
where for the whole world
50\% of the population
owns only 2\% of total wealth, while 10\% of population
owns 75\% of total wealth and 1\% of population owns 38\%
of total wealth \cite{piketty1,piketty2}.

\begin{figure}[t]
\begin{center}
\includegraphics[width=0.46\textwidth]{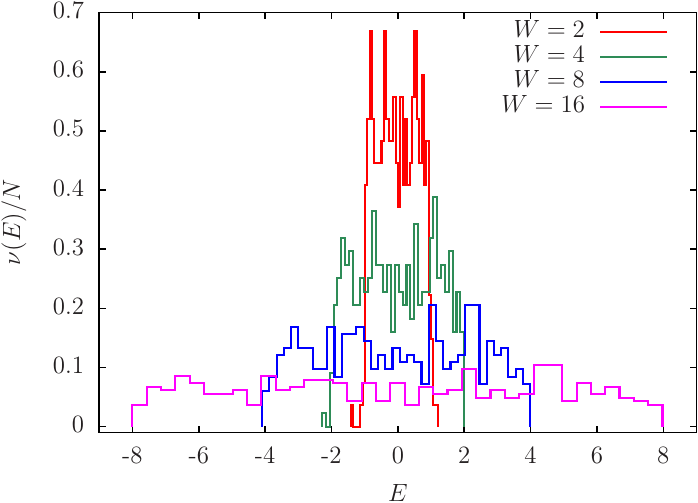}
\end{center}
\vglue -0.3cm
\caption{\label{fig1}
Rescaled density of states $\nu(E)/N$ with $\nu(E) = d m/dE_m$ 
for the energy eigenvalues $E_m$ 
of the matrix $H$ for the SSS model at $W=2,4,8,16$ 
given by (\ref{eqHdef}) for $f=0.1$ and $\kappa=0.5$. 
The normalization is 
given by $\int dE\, \nu(E)=N$. 
}
\end{figure}

We also note that statistical phenomena similar to
RJ  condensation, 
called constraint-driven condensation \cite{satya},
were studied in \cite{satya,trizac,marsili}  showing the existence of
different condensate phases for such  systems
as coalescence in granular media, jamming in traffic, 
gelation in networks \cite{satya}
and financial data analysis \cite{marsili}.
Here we discuss a new application of such type of phenomenon
for social stratification systems arguing that it
describes wealth inequality in the world. 

We mainly focus on the SSS model with Hamiltonian dynamics 
according to (\ref{eqIntegrals}), (\ref{eqNLeq1}). However, following 
the approach of \cite{ourturb}, we also consider a modified model
with pumping (absorption) of energy and norm at low (high) 
energy eigenmodes. This creates a turbulent flow from
low energy states (poor wealth layer of society)
to high energy states (rich wealth society layers)
which is in a sense similar to the KZ turbulent
energy flow from long to short wave lengths \cite{zakharovbook}.
We argue that this SSS turbulent system 
can be viewed as a simplified mathematical model
of Marxist theory for classes and wealth distribution
in a society. 

Of course, the problem of origins, structure and wealth inequality
related to a social stratification in a society
is very complex and we do not claim that we solved this problem, but we 
argue that the studied mathematical SSS model gives certain grounds
for understanding this phenomenon. 

The work is constructed as follows: the model description
is given in Section II with some additional technical details 
about RJ thermalization, numerical method and studied quantities, results 
for chaotic Hamiltonian dynamics are presented
in Section III, the results for the case with pumping and absorption
corresponding to dynamical turbulence are given in Section IV,
Lorenz curves of wealth inequality
described by the considered models are presented in Section V
and the discussion of the results is given in Section VI.

\section{Model description} 
\label{sec2}

\subsection{RJ thermalization and Entropy}

Assuming chaotic dynamics for a sufficiently large value of $\beta$, 
the Hamilonian system (\ref{eqNLeq1}) corresponds to the specific case of a 
microcanonical ensemble, with conserved energy $\mathcal{H}=E$ 
and with an additional constraint for the norm parameter $\eta=1$ which is 
also conserved. Technically, it is more convenient to treat 
this ensemble in the framework of the grand 
canonical ensemble which is justified in the limit $N\gg 1$. 
Then the quantities $\eta$ and $\mathcal{H}$ may fluctuate ``freely'' 
and only their average values are fixed by $\langle\eta\rangle=1$ and 
$\langle \mathcal{H}\rangle=E$ with some given energy $E$ value which is 
a parameter of the model. 

This particular case leads to the Rayleigh-Jeans (RJ) thermal 
distribution over 
linear eigenmodes $\phi_{n}^{(m)}$ of the matrix $H$ 
with eigenenergies $E_m$ (which are taken to be ordered 
$E_1<E_2<\ldots<E_N$). Here it is assumed that the typical contribution 
of the $\beta$-term in (\ref{eqIntegrals}) 
to the global energy is small that corresponds 
to $\xi\gg 1$ where $\xi=(\sum_n |\psi_n|^4)^{-1}$ 
(assuming normalization $\sum_n |\psi_n|^2=1$) 
is the inverse participation ratio (IPR) of the state in the initial basis. 
This 
assumption is typically well verified once we are in the thermalized regime. 
However, at initial times and depending on the initial condition 
it is possible that $\xi$ is not very large which may lead to a 
rather significant shift of the {\em linear} energy contribution (first 
term in (\ref{eqIntegrals})) between the initial and the thermalized state. 
This point will be discussed later in more detail for some 
examples. 

Using the eigenmodes $\phi_{n}^{(m)}$ of $H$, 
the wave function at time $t$
can be presented as $\psi_n(t) = \sum_m C_{m}(t) \phi_{n}^{(m)}$ 
with mode amplitudes $C_m(t)$. Then the thermalized statistical 
RJ distribution of the complex eigenmode amplitudes $C_m$ 
are independent (complex) Gaussian distributions 
with the theoretical average $\langle |C_m|^2\rangle=\rho_m = T/(E_m-\mu)$ 
\cite{landau,zakharovbook} 
where the parameters $T$ and  $\mu$ 
are the system temperature and 
chemical potential determined by the average constraints:
\begin{align}
\label{eqconstraints}
E=\mathcal{H}\approx \sum_m E_m\rho_m\quad,\quad
1=\langle\eta\rangle=\sum_m \rho_m\ . 
\end{align}
Both expressions (\ref{eqrj}) and (\ref{eqconstraints}) imply 
that $(E-\mu)/N=1/N\sum_m (E_m-\mu)\rho_m=T$ which gives a direct 
implicit equation for $\mu(E)$ using 
the 2nd equation of (\ref{eqconstraints})
\begin{align}
\label{eqImplicitmu}
1=\frac{1}{N}\sum_m \frac{E-\mu}{E_m-\mu}\ .
\end{align}
This equation provides one physically valid solution $\mu(E)$ with 
either $\mu<E_1$ (for $T>0$) or $\mu>E_N$ (for $T<0$) such that $\rho_m>0$ 
is well verified for all modes \footnote{
It is not difficult to verify that for a given spectrum 
$E_m$ the transition from $T>0$ to $T<0$ 
happens at $E=E_c=(\sum_m E_m)/N$ where $E_c$ is the ``center of mass'' 
of the energy spectrum. For the SSS model defined in Sec. II.B, we have 
$E_c\approx 0$.}. 
Once $\mu(E)$ is known (from a numerical solution 
of (\ref{eqImplicitmu})), one immediately obtains 
$T(E)=(E-\mu(E))/N$. 

We mention that the RJ distribution is a limiting case of the 
Bose-Einstein distribution, for quantum oscillators, 
at high temperatures $T \gg E_m - \mu$:
\begin{equation}
 \rho_m=\frac{1}{\exp[(E_m-\mu)/T]-1} \; ({\rm BE}) .
\label{eqbe}
\end{equation}
The steady-state RJ distribution of $\rho_m$ (\ref{eqrj})
allows also to determine the von Neumann entropy $S_q$, analogous to 
the quantum entropy, 
and the classical Boltzmann entropy $S_B$ given by the expressions 
\cite{picozziphrep,ourfiber}:
\begin{equation}
S_q = - \sum_m \rho_m \ln \rho_m\ ,\ 
S_B = \sum_m \ln(\rho_m)+S_0
\label{eqentropy}
\end{equation}
where $S_0$ is a constant which we choose arbitrarily as 
$S_0=N[1+\ln(N^2\pi)]$ to fix 
a certain numerical offset 
which ensures that typically $S_B>0$ and that $S_B/N$ and $S_q$ have 
comparable values. (See also \cite{wth} for a motivation of this choice 
using elementary phase space cells of volume $h_B^N$ with $h_B=1/N^2$.)
We mention, that the expression for $S_B$ is 
initially derived for the case of 
the thermalized situation but it can be quite easily verified \cite{wth} 
that this expression is also valid outside equilibrium if we assume 
that the amplitudes $C_m$ 
have still independent Gaussian distributions but with arbitrary averages 
$\langle |C_m|^2\rangle=\rho_m$ where $\rho_m$ are arbitrary 
parameters which are potentially different from their thermalized 
values (\ref{eqrj}). This point is important when using 
(\ref{eqentropy}) for numerically 
data at shorter time values outside equilibrium. 
For the Boltzmann entropy we have
the usual thermodynamic relation $d S_B/dE = 1/T$ 
\cite{landau,picozziphrep,ourfiber,wth} 
while the von Neumann entropy $S_q$ determines effectively a populated 
number of states at system energy $E$. 
We mention that the relation $dS_B/dE=1/T$ can also be verified directly 
by a simple calculation 
from (\ref{eqentropy}) using the thermalized expressions 
$\rho_m(E)=T/(E_m-\mu(E))$, $T(E)=(E-\mu(E))/N$ since the condition 
$\sum_m \rho_m=1$ ensures that the contributions $\sim \mu'(E)$ drop out. 
Certain properties of $S_q$ and $S_B$ are also discussed in 
\cite{ourfiber,wth}.

\subsection{Hamiltonian dynamics in the SSS models}

To model the social network links of a weighted adjacency
matrix $A_{ij}$, we choose the real network of scientific
collaboration collected by Mark Newman
\cite{newmannets,newman2001,newman2006,newman2006ref84} 
with values of $A_{ij}$ given by Eq. (2) of \cite{newman2001} and 
which are taken directly from the raw data of \cite{newman2001,newman2006}.
This nondirected  network with $A_{ij}=A_{ji}$ has $N=379$ nodes and
$N_\ell =  1828$  off-diagonal
links with a moderate average number of links per node $\chi \approx 4.82$ 
(all other matrix elements of $A_{ij}$ are zero). Therefore the 
matrix $A$ is very sparse. Links $i\to j$ and $j\to i$ for different nodes 
$j\ne i$ are counted twice in the definition 
of $N_\ell$ and self links $i\to i$ are absent. 

In the following, we choose the matrix $H$ in (\ref{eqNLeq1}) 
as 
\begin{equation}
\label{eqHdef}
H=D+f(A+\kappa H^{\rm GOE})
\end{equation}
where $A$ is the adjacency matrix introduced above 
using the data of \cite{newman2001}, 
$D$ is a diagonal matrix (see below for details) and 
$H^{\rm GOE}$ is a random GOE-matrix from the Gaussian orthogonal ensemble 
\cite{wigner,mehta},
with a semicircle density of states of radius unity (at 
large $N\gg 1$), which has 
random Gaussian matrix elements with zero mean and 
variance $\langle (H^{\rm GOE}_{n,n'})^2\rangle = (1+\delta_{n,n'})/(4(N+1))$.
Furthermore, the parameter $f=0.1$ is a simple scaling factor and $\kappa=0.5$ 
is a coefficient for the weight of $H^{\rm GOE}$. 

The matrix $H^{\rm GOE}$ is introduced 
to remove a significant number of degeneracies 
in the eigenvalue spectrum of the adjacency matrix  $A$ (which typically has 
simple fractions for the non-zero matrix elements and a quite particular 
algebraic structure). 
At $\kappa=0.5$, the effect of the degeneracies for the eigenvalues 
of $A+\kappa H^{\rm GOE}$ are ``smoothed out'' while 
the global form of the density of states is not significantly modified 
which is illustrated in Fig.~13 of \cite{wth}. Furthermore, in \cite{wth},
further details on the degeneracies of $A$ and its particular 
eigenmode structure are discussed. 

The case $D=0$, $f=1$ has been discussed in detail in \cite{wth} concerning 
the properties of $H$ and the issue of RJ thermalization in (\ref{eqNLeq1}). 
In particular, Fig.~13 of \cite{wth} shows that the eigenenergies of $H$ are 
distributed approximately in the energy interval $-6 < E_m <10$. 
The main problem of this model is that its density 
of states is highly inhomogeneous with a strong peak
in the middle of the energy band and this feature
is not typical for a society
with social stratification
corresponding to an approximately constant density of states.
To obtain a more realistic description of the model, 
we take into account the effect of social stratification by adding 
in (\ref{eqHdef}) 
the diagonal matrix $D$ with elements $D_{nn'}=D_n\,\delta_{nn'}$ 
where $D_n$ is random and uniformly distributed in the
interval $[-W/2,W/2]$. For $f=0.1$, $W\ge 2$, 
the density of states $\nu(E_m) = d m/d E_m$ of the eigenenergies 
$E_m$ of $H$ is mainly determined by the diagonal elements $D_n$ 
modeling social stratification of wealth in a society. In particular, 
$\nu(E_m)\approx N/B=  const$ where 
the energy band width is roughly $B\approx W$. Both points are 
demonstrated in Fig.~\ref{fig1} for several values $2\le W\le 16$. 

We mention, that the results presented in Fig.~\ref{fig1} and other 
subsequent figures below concern one specific random realization of 
both matrices $D$ and $H^{\rm GOE}$ in (\ref{eqHdef})
but we verified that other random realizations 
give very close results. In particular, concerning 
$H^{\rm GOE}$ the coefficient $\kappa=0.5$
is relatively weak producing essentially only a spectral smoothing of
the eigenvalues of $A$. 

\begin{figure}[t]
\begin{center}
\includegraphics[width=0.46\textwidth]{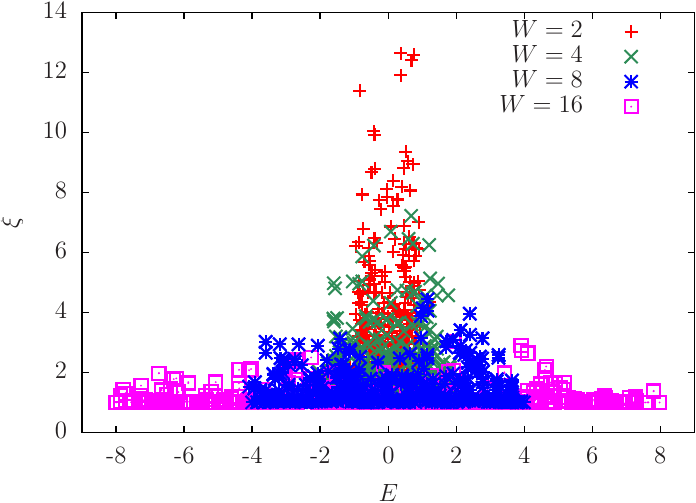}
\end{center}
\vglue -0.3cm
\caption{\label{fig2}
Energy dependence of IPR $\xi$ 
of eigenstates of the matrix $H$ for the SSS model
at $W=2, 4, 8, 16$ given by (\ref{eqHdef}) for $f=0.1$ and $\kappa=0.5$. 
}
\end{figure}

It is convenient to characterize the eigenstates $\phi_{n}^{(m)}$ of
the linear Hamiltonian $H$ by their IPR $\xi$ (see above for the 
precise definition) 
which is often used to describe the localization properties 
of eigenstates in disordered solids 
(see e.g. \cite{mirlin}) since $\xi$ corresponds 
roughly to the number of nodes over which the eigenstate is localized. 
The dependence of the IPR on the eigenstate energy $E$
is shown in Fig.~\ref{fig2}. Typical values of $\xi$ decrease with 
increasing parameter $W$ since the factor $f$ in (\ref{eqHdef}) 
is relatively small.
At the same time, the IPR seems to be randomly distributed with 
values in some interval $[1,\xi_{\rm max}]$ with the rough value 
$\xi_{\rm max}\sim 25/W$ and with smaller values at the spectral 
boundaries. 

For comparison, the RMT model used in \cite{rmtprl} corresponds to typical 
values $\xi\sim N$ where $N$ is the size of the RMT-matrix with strongly 
ergodic eigenstates. Generally, we expect that for the issue of 
thermalization in (\ref{eqNLeq1}), large $\xi$ values are favorable 
to achieve thermalization more quickly and at more modest values of 
$\beta$ (still slightly above the chaos border $\beta>\beta_{ch}$). 
Here for the SSS model typical values of $\xi$ are much smaller. 
At $W=8$, we have still many 
IPR values $\sim 2-3>1$ with some state mixing and also 
$W>f \tilde B\approx 1.6$ where $\tilde B\approx 16$ is the bandwidth of 
$A+\kappa H^{\rm GOE}$. 
This value of $W$ seems to be a reasonable 
compromise between significant mixing of states and dominant diagonal. 
Therefore, in the following, we fix $W=8$, $\kappa=0.5$, $f=0.1$  
in (\ref{eqHdef}) which corresponds to a homogeneous stratification
of energy and eigenstates. In our opinion, these parameters 
correspond to a typical social stratification in a society. 

\begin{figure}[t]
\begin{center}
\includegraphics[width=0.46\textwidth]{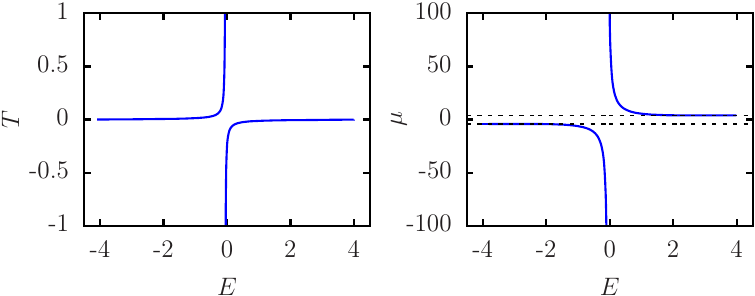}
\end{center}
\vglue -0.3cm
\caption{\label{fig3}
The left (right) panel shows the temperature $T$ 
(the chemical potential $\mu$) versus energy $E$ 
for the SSS model and the RJ case at $W= 8$, $f=0.1$, $\kappa=0.5$, $N=379$.
The dashed black lines in the right panel correspond to the values of 
$E_1=-4.09$ and $E_N=3.99$ (minimal and maximal energies $E_m$ for 
the used random realization of $H$) 
showing that either $\mu<E_1$ (for $T>0$) 
or $\mu>E_N$ (for $T<0$). The transition from $T<0$ to $T>0$ happens at 
$E=E_c=-0.053$ for the given random realization of $H$ used 
to compute the spectrum $E_m$. 
}
\end{figure}

As illustration, we show in Fig.~\ref{fig3} the energy dependence 
of the temperature $T(E)$ and the chemical potential $\mu(E)$ 
for RJ thermalization obtained from the numerical solution of 
(\ref{eqImplicitmu}) 
together with the relation $T(E)=(E-\mu(E))/N$ for the eigenvalue 
spectrum of (\ref{eqHdef}) for these parameters. Depending on 
$E<E_c$ ($E>E_c$), we have positive (negative) temperatures $T>0$ ($T<0$) 
and $\mu<E_1$ ($\mu>E_N$) where $E_c=(\sum_m E_m)/N$ is the 
center of mass for the energy spectrum. 
For our random realization of $H$ we have $E_c=-0.053\approx 0$. 
In Section III, we will present 
evidence for the presence of thermalization of typical states at sufficiently 
large values of $\beta$ (and the case $T>0$). 

In this work, we solve numerically the dynamical system (\ref{eqNLeq1}) 
for the SSS model for the matrix $H$ given by (\ref{eqHdef}) 
using a symplectic 4th order integrator (see \cite{rmtprl} and references 
therein). The advantage of this method is that it conserves exactly 
(up to usual numerical rounding errors) the quantity $\eta$ while the 
energy $\mathcal{H}$ is only approximately conserved. 
The numerical time integration step is chosen such that the numerical 
energy fluctuations of $\mathcal{H}$ are well below $10^{-3}$. 

Usually, we start the  time evolution (\ref{eqNLeq1})
with an initial state $\psi(t=0) = \phi_{n}^{(m_0)}$ located at the eigenmode
with energy $E_{m_0}$, i.e. $C_m(t=0)=\delta_{m m_0}$ ($=1$ for $m=m_0$ 
and $=0$ for $m\neq m_0$). 
Using the numerically obtained values of $C_m(t)$, we compute 
the mode probabilities $\rho_m$ 
as the time average $\rho_m = \langle|C_m(t)|^2\rangle$ 
over successive time intervals $2^{l-1}\le t<2^{l}$ 
for discrete integer values 
of $l=1,2,3,\ldots,l_{\rm max}$ with possible maximal values 
$l_{\rm max}=30$ (i.e.~$t_{\rm max}=2^{30}$). These numerical $\rho_m$ 
values, for a given time interval, can for example be used to 
compute numerical 
entropy values for both $S_q$ and $S_B$ which can be compared with their 
theoretical thermalized values (the results are presented in the next 
Section). 

At exactly $t=0$, we have 
formally $S_q=0$ and $S_B \to -\infty$ since 
$\rho_m(t=0)=|C_m(t=0)|^2=0$ for $m\neq m_0$. However, due to 
the used time average procedure to compute $\rho_m$ from 
$\langle |C_m(t)|^2\rangle$ over successive intervals $2^{l-1}\le t<2^l$ 
the initial divergent value of $S_B$ is not visible. In particular, 
the first used interval corresponds to $l=1$ such that $1\le t<2$ and 
the very initial time values $0\le t<1$ are simply not 
used in our $\rho_m$ data from which $S_q$ 
and $S_B$ are extracted by (\ref{eqentropy}). Initial values $S_B(t=2)$ 
obtained for the initial interval may be negative but typically they 
become rather quickly positive with increasing $t$, 
which is also due to our choice of 
the constant $S_0$ used in (\ref{eqentropy}). 

In this work, we consider the cases with nonlinear parameter $\beta \ge 2$
when the dynamics is clearly chaotic. We do not try to determine 
the precise value of the chaos border $\beta_{ch}$ of this model
since some discussion is given in \cite{rmtprl}. 
Furthermore, the actual determination of the chaos
border in a nonlinear system with many degrees of freedom is a 
complicated problem due to Arnold diffusion and
other subtle nonlinear effects \cite{lichtenberg,chirikov1979}.

\subsection{Dissipative dynamics and turbulence in SSS}

In \cite{ourturb} a particular 
modification of the dynamical system (\ref{eqNLeq1}) 
(for the case of a random matrix $H=H^{\rm GOE}$) 
was proposed to model Kolmogorov-Zakharov turbulence. The idea is 
to incorporate in the system norm and energy pumping at low energy scales and 
absorption at high energies. 
For a pumping strength above
a certain chaos border, a global chaotic attractor
appears with a stationary energy flow through the initial 
energy interval \cite{zakharovbook}.
In this regime, the steady-state norm distribution
is described by an algebraic decay
with an exponent in agreement with the KZ theory.
This distribution is close to the RJ one (\ref{eqrj}).
Below the chaos border there is a KAM regime
where the cascade flow is broken.

For the SSS model, as in \cite{ourturb}, we rewrite Eq. (\ref{eqNLeq1}) for 
the amplitudes $C_m(t)$ in the energy eigenbasis and 
add a new term for pumping and absorption \cite{ourturb}:
\begin{align}
\label{eqturb}
i {{\partial C_m} \over {\partial {t}}}
=& E_m C_m + i (\gamma_m - \sigma_m \vert C_m \vert^2)  C_m\\
\nonumber
&+ \beta \sum_{{m_1}{m_2}{m_3}}
V_{{m}{m_1}{m_2}{m_3}}
C_{m_1}C^*_{m_2}C_{m_3}\quad,\\
\label{eqtransV}
V_{{m}{m_1}{m_2}{m_3}}=& \sum_{n}\phi_n^{(m)*}\phi_n^{(m_1)}
\phi_n^{(m_2)*}\phi_n^{(m_3)}\ .
\end{align}
For $\gamma_m=\sigma_m=0$ (\ref{eqturb}) is exactly equivalent 
to (\ref{eqNLeq1}) (taking into account 
the linear transformation $\psi_n\to C_m$). 
In (\ref{eqturb}) the transitions between linear eigenmodes
appear only due to the nonlinear $\beta$-term
with the transition matrix elements given by (\ref{eqtransV}) 
in terms of the eigenmodes $\phi_n^{(m)}$. For the RMT case for $H$, as 
in \cite{rmtprl,ourturb}, they have typical values 
$V_{{m}{m_1}{m_2}{m_3}}\sim N^{-3/2}$ which are typically very small 
but all modes are coupled by them and there are many contributions of them 
in (\ref{eqturb}).

For the SSS model, corresponding to $H$ given by (\ref{eqHdef}), the 
situation is more complicated due to the reduced IPR values of the eigenmodes. 
If all four modes $m,m_1,m_2,m_3$ are roughly localized on the same nodes 
(of typical number $\sim\xi$), we have rather large (``maximal'') 
transition matrix elements  
$V_{{m}{m_1}{m_2}{m_3}}\sim \xi^{-3/2}$ but this value 
may be strongly reduced if the modes are localized 
on different nodes. Globally, this gives a very complicated dynamical system 
even for the Hamiltonian case $\gamma_m=\sigma_m=0$. 

\begin{figure}[t]
\begin{center}
  \includegraphics[width=0.46\textwidth]{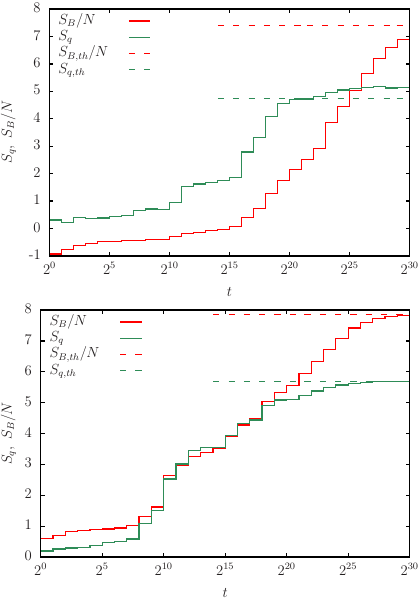}
\end{center}
\vglue -0.3cm
\caption{\label{fig4}
Time dependence of von Neumann entropy $S_q(t)$ (full red line) 
and (rescaled) Boltzmann entropy $S_B/N$ (full green line) for the SSS model;
here $W=8$, $N=379$, $\beta=2$ (top)
and $\beta=4$ (bottom); initial state is the linear
eigenmode at $m_0=20$, i.e. $C_m(t=0)=\delta_{m,m_0}$. 
Entropy values are computed from (\ref{eqentropy}) 
using time averaged $\rho_m=\langle |C_m(t)|^2\rangle$ values 
for successive time intervals $2^{l-1}\le t<2^{l}$ for 
$l=1,2,3,\ldots,30$. 
The dashed lines show the theoretical thermalized values, using 
(\ref{eqentropy}) and the RJ values for $\rho_m$ given by (\ref{eqrj}). 
}
\end{figure}

Due to the presence of the stratification term in the SSS and KZ models 
the eigenstates are localized on a relatively small  number of nodes
given by the IPR $\xi \ll N$. 
Thus the transition rate $\Gamma_\beta$ between such states,
induced by  matrix elements $V$ and the nonlinearity term $\sim\beta$,
can be estmated via the Fermi golden rule
as $\Gamma_\beta \sim V^2 \rho \sim (\beta^2 \eta^2/\xi^3) \xi^2 \sim 10$
(for parameters of Fig.~\ref{fig4} botton panel
with $\xi \sim 2$, $\eta=1$).
Since the IPR values $\xi \ll N$ are small
the spearing over the stratification energy
lattice  goes like a diffusion process with
$(\Delta m)^2 \sim \Gamma_\beta t$.
Thus the relaxation time $t_{ss}$ to the steady-state
with $\Delta m \sim N=379$ can be estimated as
$t_{ss} \sim N^2/\Gamma_\beta \sim 10^4$.
Indeed, the data of Fig.~\ref{fig4}
shows that the diffusive spreading
takes place approximately during the time interval
$2^7 \leq t \leq 2^{18}$
with $\Delta m = \exp(S_q) \sim t^{b}$
with the fit exponent $b= 0.548 \pm 0.034$
being close to the diffusion exponent $b=0.5$.
Due to small values of $\xi \sim 2$ it takes a certain time
$t \sim 2^7$ for the diffusive spreading to start. 
Thus the time scale to reach thermalization to the steady-state
is rather long. It decreases for the KZ case
when $\eta$  increases (see Figs. below in Section IV).

Using non-zero values of $\gamma_m>0$, $\sigma_m>0$ in (\ref{eqturb}) we 
can model either pumping of modes or for $\gamma_m<0$, $\sigma_m=0$ 
dissipation of modes. In this case, the quantities $\eta$ and 
$\mathcal{H}$ are not conserved and the dynamical system (\ref{eqturb}) 
is no longer Hamiltonian. The positive value $\sigma_m>0$ (if 
$\gamma_m>0$) implies that amplitudes of the corresponding modes 
typically saturate to $|C_m(t)|\to C_{\rm sat}=\sqrt{\gamma_m/\sigma_m}$ 
at long times \cite{ourturb}. 

To model the energy flow of a direct cascade
from low to high energy modes $m$ we use pumping 
$\gamma_m=\gamma>0$ for the 4 lowest energy modes at $m=1,2,3,4$ 
with corresponding saturation coefficients
$\sigma_m = \sigma>0$ and dissipation $\gamma_m=- \gamma<0, \sigma_m=0$
for the 4 highest energy modes with $m=N, N-1, N=2, N-3$. 
For all other $m$ values we take $\gamma_m=\sigma_m=0$.
Here $\gamma$ and $\sigma$ are two parameters of the model 
and in most cases, as in \cite{ourturb}, we choose $\gamma=\sigma=0.01$ 
with saturation amplitude $C_{\rm sat}=1$ (for modes with $m\le 4$). 

For the case of an RMT matrix $H=H^{(\rm GOE)}$, the 
properties of the KZ direct cascade have been discussed in 
detail in \cite{ourturb}. 
In this work here, we apply the same approach, based on 
Eq. (\ref{eqHdef}) for $H$, to model the direct
cascade of turbulent flow  from low to high energies (or wealth  values)
in the SSS model.
In a certain sense, this SSS turbulence system
can be considered as a mathematical model
of Marxist theory \cite{marx} when wealth
is created by workers in the lowest wealth layer
of society and it propagates to the layer with highest wealth
being absorbed there. 

As initial condition, we choose as in \cite{ourturb}, random uniform 
values for the amplitudes $C_m(t=0)$ with $1\le m\le 8$ such that the 
initial norm is $\eta(t=0)=0.01$ and other initial amplitudes are 
$C_m(t=0)=0$ (for $m>8$). Furthermore, we use the same numerical method 
as in \cite{ourturb} which is an adapted version of the symplectic integrator 
used in \cite{rmtprl}. We discuss the results and the properties
of the steady-state distribution $\rho_m$
in the SSS turbulence model in Section IV. Here, we mention briefly 
that the initial norm $\eta(t=0)=0.01$ quickly increases and saturates 
at long times at values $\eta(t)\to\eta_{\rm sat}\gg 1$. 
The numerical values $\rho_m$ for the KZ case are obtained by the same 
time average $\rho_m=\langle |C_m(t)|^2\rangle$ over intervals 
$t\in[2^{l-1},2^l\,]$, $l=1,2,\ldots$. However, here $\eta=\sum_m \rho_m$ 
is no longer conserved. 

\section{Results for Hamiltonian dynamics of the SSS model}
\label{sec3}

In this Section, we present and discuss the numerical results 
for the Hamiltonian system of the SSS model obtained by solving 
(\ref{eqNLeq1}) with the initial condition $\psi_n(t=0)=\phi_n^{(m_0)}$
(or equivalently $C_m(t=0)=\delta_{mm_0}$), i.e. the initial state is just 
the eigenmode $\phi^{(m_0)}$ of $H$. 
To obtain thermalization within numerically accessible time scales, 
we need to take a relatively strong
nonlinear coefficient $\beta$
since the relaxation rate to the steady-state
is expected to be proportional to $\Gamma \propto \beta^2$
(as for the Fermi golden rule)
such that the relaxation time behaves as $t_\Gamma \sim 1/\Gamma$. 
For this reason, we focus here on the two cases $\beta=2$ 
and $\beta=4$. 

In order to establish the presence of RJ thermalization, 
we first compute from the numerical data both 
entropy quantities given in (\ref{eqentropy}). 
Their time dependence is shown in Fig.~\ref{fig4} for the initial mode 
$m_0=20$ and both values of $\beta$. 
Both entropy quantities increase with time and reach saturation 
at the theoretical values obtained from (\ref{eqentropy}) using 
the theoretical RJ thermalized $\rho_m$ values given in (\ref{eqrj}). 
It seems that at $\beta=2$ the 
thermalization is not fully achieved at the largest time value $t=2^{30}$ 
while for $\beta=4$ the matching with the thermalized entropy values at 
the largest time is very good. 

\begin{figure}[t]
\begin{center}
  \includegraphics[width=0.46\textwidth]{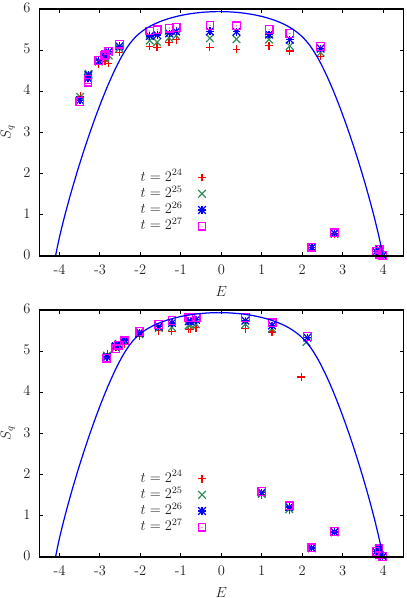}
\end{center}
\vglue -0.3cm
\caption{\label{fig5}
Dependence of von Neumann entropy $S_q$ on the average linear energy 
$E=\sum_m E_m\rho_m\approx {\cal H}$
for several initial eigenstates for different $m_0$ and $t$ values
for $\beta=2$ (top) and $\beta=4$ (bottom);
the other parameters are $W=8$, $f=0.1$, $\kappa=0.5$ and $N=379$. 
The blue line shows the theoretical thermalized value of $S_q$. 
As in Fig.~\ref{fig4}, $S_q$ is computed from (\ref{eqentropy}) using 
either numerically averaged $\rho_m$ values (data points) or 
the thermalized RJ values (\ref{eqrj}) (blue line). 
}
\end{figure}

In Fig.~\ref{fig5}, we show the energy dependence of the 
von Neumann entropy $S_q$. The blue curve corresponds to the theoretical 
thermalized energy dependence (obtained by combining (\ref{eqentropy}) 
and (\ref{eqrj})). The data points correspond to a selection 
of 23 states for different initial modes, including the 4 bottom 
($m_0=1,2,3,4$), the 4 top ($m_0=N-3,N-2,N-1,N$) 
$m_0$ values and also the case $m_0=20$ (used in 
Fig.~\ref{fig4}) for 
time values $2^{24}\le t\le 2^{27}$ (more precisely $S_q$ 
is obtained from (\ref{eqentropy}) with 
$\rho_m(t)$ being the average $\langle |C_m(\tau)|^2\rangle$ 
for $\tau\in[t/2,t]$). 

Before discussing the results shown in Fig.~\ref{fig5}, 
we mention that for the data points the energy value $E$ on the abscissa 
corresponds to the average linear system 
energy $E=\sum_m E_m \rho_m$ (Note that here 
and in the following we  use the notation $E$ only for the linear 
energy $E=\sum_m E_m\rho_m$). The value of the 
linear energy $E$ may be quite different from 
the energy $E_{m_0}$ of the initial mode due a significant energy 
shift effect caused by a quite large value of $\beta$ and 
small initial IPR value $\xi$ of the initial state. To see this 
more clearly, we write the global conserved energy $\mathcal{H}$ of 
(\ref{eqIntegrals}) at two time values $t=0$ and $t=t_f$:
\begin{align}
\label{eqEnshift}
\mathcal{H}&=E_{m_0}+\frac{\beta}{2\xi(t=0)}=
\sum_n E_m\rho_m(t_f)+\frac{\beta}{2\xi(t_f)}
\end{align}
where $\xi(t)$ is the IPR of the state $\psi(t)$ at time $t$ (in the 
initial node basis) and $t_f$ is some 
long time value such as $t_f=2^{27}$ used in Fig.~\ref{fig5}. At such 
times, and assuming thermalization, typical values of the 
IPR (of $\psi(t_f)$) are rather large $\xi(t_f)\gg 1$ such 
that the energy contribution of the non-linear term can be neglected 
and we have $\mathcal{H}\approx E= \sum_m E_m\rho_m(t_f)$. 
However, at $t=0$, the initial state $\psi(t=0)$, being the eigenmode 
$\phi^{(m_0)}$, has a rather small IPR value, typically below 3-4 
(see Fig.~\ref{fig2}), such that 
$\Delta E=\beta/(2\xi(t=0)\approx E-E_{m_0}$ takes significant values being 
around 10-20\% of the bandwidth $B\approx 8$. 
For example, for $\beta=4$ and $m_0=1$, 
we have $\xi(t=0)=1.52$, $E_{1}=-4.09$, 
$E=\sum_m \rho(t_f)\rho(t_f)=-2.83$ with $\Delta E=1.27$ which 
is close to $\beta/(2\xi(t=0)=1.32$ (the remaining difference of 
0.05 is due to the remaining small value of $\beta/(2\xi(t_f))\ll 1$ 
in (\ref{eqEnshift}); residual 
technical fluctuations in the value of $\mathcal{H}$ 
due to the numerical method are clearly below $10^{-3}$). 

This energy shift effect from $E_{m_0}\to E=E_{m_0}+\Delta E$ 
affects nearly all data points 
in Fig.~\ref{fig5}. For $\beta=2$ the linear energies 
are in the interval $-3.5<E<4$ and for $\beta=4$ they are even in the 
interval $-3<E<4$ despite the fact that $E_1=-4.09$ (which corresponds to 
the first data point with minimal $E$ in both panels of Fig.~\ref{fig5}). 
Since $\beta>0$ this effect is favorable to achieve thermalization 
for modes with $E_{m_0}<0$. For example for $m_0=1$ and $\beta=4$ 
the initial linear energy 
is shifted from $-4.09$ to $-2.83$ and the latter provides 
a significant gap from the 
left spectral border which facilitates thermalization. 
The numerical values of $S_q$ for $E<0$ and a few data points 
at $E>0$ shown in Fig.~\ref{fig5}
approach with increasing time $t$ indeed the theoretical RJ curve even 
though at $t=2^{27}$ thermalization is not yet perfect, in particular 
for $\beta=2$. Some modes with small $m_0$ value have actually values 
of $S_q$ above the theoretical curve. In \cite{rmtprl}, it was
also observed that $S_q$ (for a mode with small $m_0$ value) is above the 
theoretical curve at intermediate time scales but at longer time scales 
it finally reaches the theoretical value 
(see SupMat Fig.~S1 of \cite{rmtprl}). We expect a similar behavior 
for the small $m_0$ modes in Fig.~\ref{fig5} where we have still 
$S_q$ above the theoretical RJ curve for $t=2^{27}$ but here the 
necessary time scales to reach the RJ curve are numerically 
not accessible. 

However, for modes with $E_{m_0}>0$ thermalization becomes 
more difficult due to the energy shift effect. 
There are two cases like this: the first one concerns  
the 4 top modes where $E_{m_0}\approx E_N=3.98$.
For these states it is impossible to have 
an energy shift effect since we have mathematically 
$E=\sum_m E_m\rho_m\le E_N$ such that 
$\Delta E=E-E_{m_0}\approx 0$. Therefore, 
according to (\ref{eqEnshift}), 
the IPR $\xi(t_f)$ must be close to its initial value $\xi(t=0)$ which 
is actually $\xi(t=0)\approx 1$ for these 4 top modes (see also 
Fig.~\ref{fig2}) and delocalization (thermalization) is simply 
impossible due to the conservation of $\mathcal{H}$. (Note that for 
negative $\beta<0$ we would have such an effect at the 
lower spectral boundary.)

The second case of ``difficult modes'' concerns {\em some states} 
with small $S_q$ values and 
with energies $E$ above a certain threshold $E_{th}$ 
(with $E_{th}\approx 2.2$ for $\beta=2$ and 
$E_{th}\approx 1$ for $\beta=4$) but still with a significant gap 
with respect to the right spectral border $E_N=3.98$. For these states, 
the $S_q$ values also remain 
far from the RJ theory, clearly with $S_q>0$ but also with 
$S_q<0.6$ ($S_q<2$) for $\beta=2$ ($\beta=4$). For such states, it 
is not impossible to have an energy shift effect. However, the data 
shows that for them we typically also have $E\approx E_{m_0}$ (at 
$t=2^{27}$) resulting in a small IPR value and absence of thermalization. 
However, these states can still 
profit from the energy shift effect and it is indeed possible that 
thermalization for them sets in at much longer time scales than visible 
in Fig.~\ref{fig5} and numerically accessible. 

There are also some (rather) well thermalized {\em other} modes with 
$E>E_{th}$ but closer inspection shows that for them we have 
$E_{m_0}<E_{th}$ and they have a strong energy shift effect (which 
allows the reduction of the IPR).

We also remind that the adjacency matrix $A$ has 
significant degeneracies in its spectrum 
(see the detailed discussion in \cite{wth})
that still plays a role even when the GOE $\kappa$-term is added
in (\ref{eqHdef}). Furthermore, the matrix $A$ is very sparse 
with a relatively small
number of links per node $\chi \approx 4.82$
(for comparison other networks such as Wikipedia have $\chi \approx 20$). 
We think that both points are responsible for the observation 
that for the states with 
$S_q\approx 1$-$2$ (those with $E\approx E_{m_0}>E_{th}$) 
thermalization is  very slow. 

We note that similar type of deviations 
at high energies $E$ have been seen in \cite{wth}
without the stratification term $D$ (see e.g. Fig.~14 in \cite{wth}).

We do not show a figure for the energy dependence of $S_B$ but 
the results are globally 
similar to Fig.~\ref{fig5} for $S_q$.

\begin{figure}[t]
\begin{center}
  \includegraphics[width=0.46\textwidth]{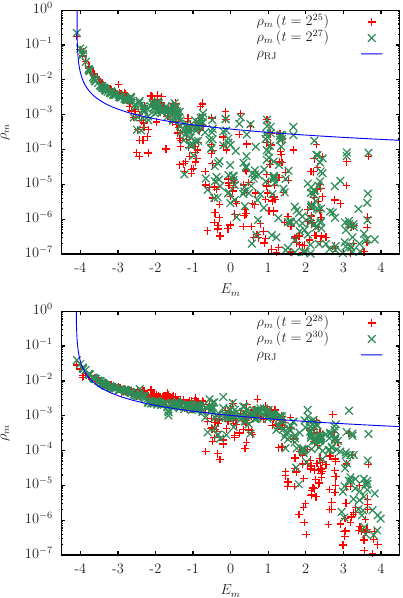}
\end{center}
\vglue -0.3cm
\caption{\label{fig6}
  Dependence of average eigenstate probability $\rho_m$ 
  on eigenstate energy  $E_m$ for the initial states 
  at $m_0=1$ (top) and $m_0=20$ (bottom) at two different time values; 
  the energy values are $E_{m_0}=-4.09$, $E=\sum_m E_m\rho_m=-3.49$ 
  ($m_0=1$, top) and $E_{m_0}=-3.48$, $E=-2.52$ 
  ($m_0=20$, bottom); 
  the blue curve shows the theoretical RJ distribution (\ref{eqrj}) 
  using the given value of $E$ for each mode; 
  other parameters are $\beta =2$, $W=8$, $f=0.1$, $\kappa=0.5$ and $N=379$. 
}
\end{figure}

\begin{figure}[t]
\begin{center}
  \includegraphics[width=0.46\textwidth]{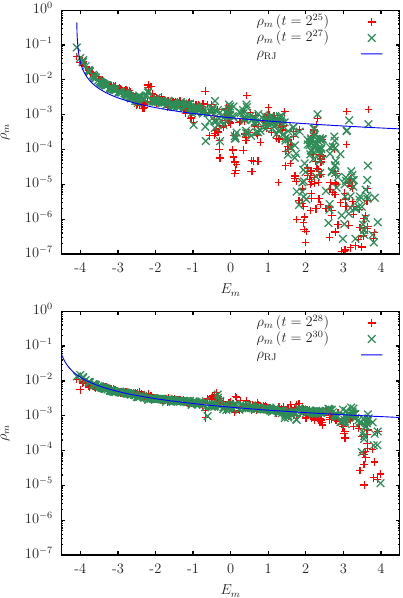}
\end{center}
\vglue -0.3cm
\caption{\label{fig7}
  Same as Fig.~\ref{fig5} but for $\beta=4$ 
  and energy values $E_{m_0}=-4.09$, $E=-2.83$ 
  ($m_0=1$, top) and $E_{m_0}=-3.48$, $E=-1.55$ 
  ($m_0=20$, bottom). 
}
\end{figure}

Examples of several probability distributions $\rho_m$
obtained at large times are shown in Fig.~\ref{fig6}
for $\beta=2$ and Fig.~\ref{fig7} for $\beta=4$ 
for the two initial modes $m_0=1,20$. 
The results show that with an increase of
total evolution  time  the values of $\rho_m$
approach the theoretical RJ distribution (\ref{eqrj}).
This process is more rapid for $\beta=4$ than for $\beta =2$ 
since the dynamical thermalization
takes place only due to the nonlinearity
with the relaxation rate $\Gamma \propto \beta^2$.
The thermalization takes a long time to
reach states at high energies but
we expect that the thermalization will eventually
reach these states at  longer times.
We note that a similar slow relaxation to the thermal state
at high energies is also seen in the NLIRM model
with diagonal matrix elements
that approximately corresponds to the stratification term
in the SSS model (see SupMat Fig.~S14 in \cite{rmtprl}).

We also note that in Figs.~\ref{fig6},~\ref{fig7}
the RJ condensation fraction
is relatively weak being e.g. about 25\% of total norm
in the linear groundstate eigenmode
(see top panel in Fig.~\ref{fig6}). This is essentially due to 
the energy shift effect. For example, for $m_0=1$ and $\beta=4$, 
we have $E_{m_0}=E_1=-4.09$ while for the RJ thermalization curve 
we need to use $E=\sum_m E_m\rho_m=-2.83$. 
In other systems the condensation fraction
can be around 80-90\% of total norm
as e.g. for the NSE dynamics in quantum chaos fibers \cite{ourfiber}
or in the social networks without stratification term ($D=0$ in (\ref{eqHdef}))
\cite{wth}.

Globally, the data presented in this Section strongly 
indicate the presence of RJ thermalization for the SSS model 
despite the sparse matrix structure of $A$ and the strong diagonal 
with $W=8$, which both tend to reduce typical IPR values of eigenstates. 
However, for this model it is more difficult to access the necessary 
numerical time scales and deviations for some modes with largest energies 
can be explained by the (absence of the) energy shift effect which is due 
to large $\beta$ and small initial IPR values. 
Even though the adjacency matrix $A$
has certain specific features (degeneracies, sparsity) we argue that it is
important to use the data of a real social network of
scientific collaboration \cite{newman2006}
to study the properties of
social stratification of society.

\section{Results for KZ like turbulence in SSS models}
\label{sec4}

In this Section we present results
for the dynamical KZ turbulence in the SSS model
described above in the Section II.C.
In this model there is norm and energy
pumping on the lowest $m=1,2,3,4$ states and
absorption on highest states at $m=N-3, N-2, M-1, N$. 
The initial condition at $t=0$, corresponds to random 
uniform amplitudes $C_m(t=0)$ for $1\le m\le 8$ (and 
$C_m(t=0)=0$ for $m>8$) with 
initial norm $\eta(t=0)=\sum_m |C_m(t=0)|^2=0.01$. 
A similar type of model of KZ turbulence (RMT of KZT)
was studied in \cite{ourturb} for the case of an RMT matrix $H$.

Here we discuss only the case of a direct cascade
when energy pumping takes place
at low modes $m =1,2,3,4$
(analogous to long waves in turbulence) and it is absorbed at
high energy modes $m=N,N-1,N-2,N-3$
(analogous to short waves). As in the RMT case \cite{ourturb}, for 
the inverse cascade the SSS system rapidly enters a strongly nonlinear 
regime where the weak
turbulence description is invalid. Also for the SSS model
pumping at low $m$ values can be viewed as a new wealth created
by poor workers which propagates to high
wealth levels of society being absorbed
at rich oligarchic phase at highest $m $ values.
The inverse process would look rather strange for
a wealth creation and propagation in a society and we do not discuss it here.

\begin{figure}[h]
\begin{center}
\includegraphics[width=0.46\textwidth]{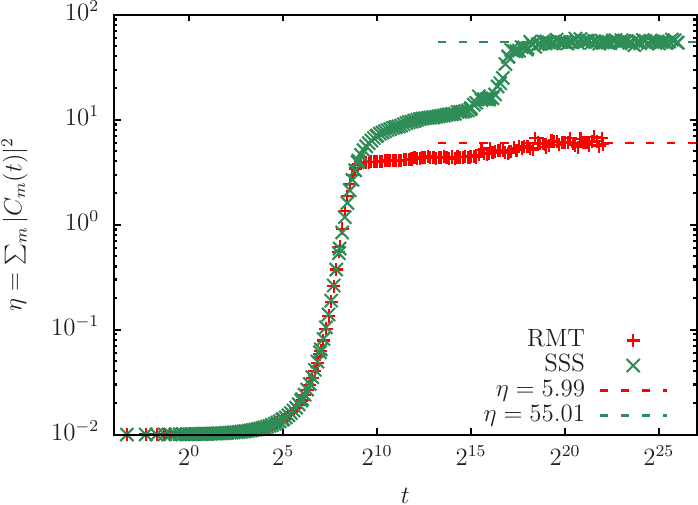}
\end{center}
\caption{\label{fig8}
Time dependence of the
norm $\eta(t)=\sum_m |C_m(t)|^2$  (non-averaged) at selected time values 
for the SSS model ($\beta=4$, $N=379$) and RMT of KZT ($\beta=1$, $N=512$) 
from \cite{ourturb}. The norm values
$\eta_{\rm av}=\langle \eta(\tau)\rangle$ with average over the 
maximal time interval $t/2\le \tau\le t$ are 
$\eta_{\rm av}=5.99$ (RMT of KZT, $t=2^{22}$, $\beta =1$) and 
$\eta_{\rm av}=55.01$ (SSS, $t=2^{26}$, $\beta=4$) (both indicated by dashed lines).
Here $\gamma=\sigma=0.01$ in the evolution of Eq.~(\ref{eqturb}) 
with pumping, absorption, initial condition as described in Sec. II.C. 
}
\end{figure}

The pumping leads to a growth of total norm $\eta$
as it is shown in Fig.~\ref{fig8} for the SSS model (at $\beta=4$, $N=379$) 
and the RMT case taken from \cite{ourturb} (at $\beta=1$, $N=512$). 
After a certain time $t_{st}$ a chaotic spreading of norm $\eta$
over linear states reaches the states with absorption
and the norm growth is stopped and the system enters in
the steady-state regime. For both models, we have roughly 
$t_{st} \sim 2^{18}$. However, for the RMT model we have
at $t \sim 2^9$ an intermediate plateau only slightly below 
the final value while for the SSS model there is an additional 
increase of $\eta$ for $2^{17}<t<2^{19}$. 
We attribute this 
to the fact that the factor $f=0.1$ in (\ref{eqHdef})
is small and thus the linear links couple diagonal
energy states in a relatively narrow energy band while
for the RMT case in \cite{ourturb} such matrix transition couple
all states inside the  energy band. 

\begin{figure}[h]
\begin{center}
\includegraphics[width=0.46\textwidth]{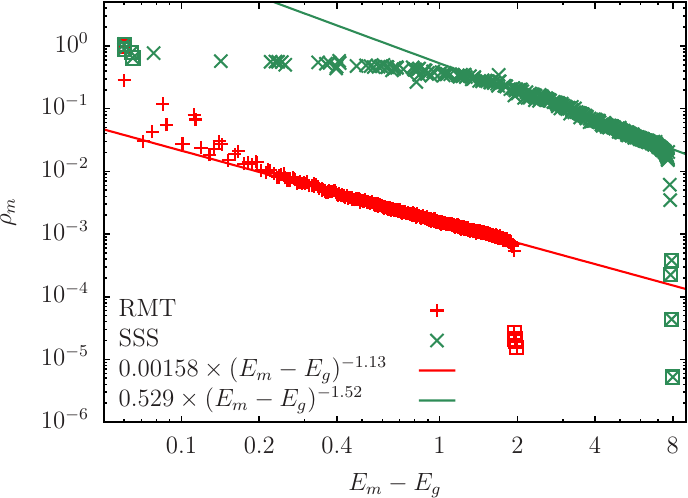}
\end{center}
\caption{\label{fig9}
Distribution $\rho_m$ versus $E_m-E_g$ in a double logarithmic representation 
for the model of KZ turbulence of \cite{ourturb} with 
$\gamma=\sigma=0.01$, pumping 
at $m=1,\ldots, 4$ and absorption at $m=N-3,\ldots, N$ (data points with 
boxed symbols); see text for the initial condition. 
Data points with $E_m-E_g<0.06$ have been 
artificially moved to $E_m-E_g=0.06$. 
The values of $\rho_m$ have been obtained from the time average 
$\rho_m(t)=\langle |C_m(\tau)|^2\rangle$ for $t/2\le \tau\le t$. 
Red $+$ symbols 
correspond to the RMT model of \cite{ourturb} for $N=512$, $\beta=1$, 
$t=2^{22}$, $E_g=-1$ (same 
data as in Fig.~3 of \cite{ourturb}) and green $\times$ symbols 
correspond to the SSS model introduced here for $N=379$, $\beta=4$, 
$t=2^{26}$, $E_g=-4$. 
The straight lines show the power law fit $\rho_m=A\,(1+E_m)^{-s_0}$ 
with $A=(1.584\pm 0.003)\times 10^{-3}$, $s_0=1.130\pm 0.004$, 
fit range $0.2\le E_m-E_g\le 1.8$ (for RMT) and 
with $A=0.5288 \pm 0.008$, $s_0=1.52 \pm 0.01$, 
fit range $1\le E_m-E_g\le 7$ (for SSS).
The values of the norm 
$\eta_{\rm av}=\sum_m\rho_m$ are $\eta_{\rm av}=5.99$ (RMT) and 
$\eta=55.01$ (SSS). 
}
\end{figure}

In Fig.~\ref{fig9}, we show the energy dependence of the steady-state
norm distribution $\rho_m$ obtained in the regime of KZ turbulence
for the SSS model ($t=2^{26}$) and the RMT model ($t=2^{22}$).
The time evolution of norm $\eta(t)$ for these two cases
is shown in Fig.~\ref{fig8}. At high energies
both distributions demonstrate an approximate
algebraic decay $\rho_m \propto 1/(E_m-E_g)^{s_0}$
with the exponents $s_0 =1.130 \pm 0.004$
for the RMT case from \cite{ourturb}
and $s_0 = 1.52 \pm 0.01$ for the SSS case. Here $E_g$ is the 
approximate spectral lower boundary ($E_g=-4$ for SSS and $E_g=-1$ for 
RMT). 
For the RMT case the exponent $s_0$ is close to the theoretical
value $s_0=1$ for the KZ turbulence \cite{zakharovbook,ourturb}
(see Eq.(3.1.10a) in \cite{zakharovbook},
also Eq.(5) in \cite{ourturb}).
For the SSS model we have $s_0 =1.52$ being
noticeably higher than the theoretical value $s_0=1$
obtained theoretically
for acoustic waves with linear spectrum $\omega_k =c k$
and 4-waves nonlinear interaction \cite{zakharovbook}.
We attribute this deviation to a smaller system size
(smaller $N$ cases have higher $s_0$ values as it is seen for the RMT model
cases in Fig.~1 of \cite{rmtprl}). 
Also in the SSS model the links of the matrix $A$ in (\ref{eqHdef})
are rather local in energy, due to the small scaling factor $f=0.1$ 
while in the RMT case and wave turbulence \cite{zakharovbook}
the couplings have a significantly broader range.
Furthermore, the spectrum $E_m$ of the SSS model is only approximately 
linear in mode number $m$
since there are random fluctuations of eigenenergies of the linear
system (\ref{eqHdef}). We argue that these particular features 
of the SSS model are responsible for the difference 
between the values of the theoretical and numerical exponents $s_0$.
Also we should note that the acoustic turbulence
with a purely linear wave spectrum without dispersion
usually has three-wave interactions \cite{kuznetsov2,kuznetsov3}
while in our model
we have the case of four-wave interactions
and a randomly fluctuating spectrum of linear modes
(thus their spectrum is only approximately linear).

Fig.~\ref{fig9} shows results for $\gamma=0.01$ for both models but 
we checked that for $\gamma=0.005$ the results give rather similar 
algebraic like distributions $\rho_m$ (apart from the total norm value).

\section{Lorenz curves for wealth inequality}
\label{sec5}

In Sections I, II it is pointed out that the
eigeneneries $E_m$ of the linear
Hamiltonian at $\beta=0$ (\ref{eqHdef})
can be also considered as individual
wealth levels (layers or states) 
of a society with social stratification.
In this framework, we can interpret the RJ thermal
distribution (\ref{eqrj}) as wealth distribution 
for a population fraction $\rho_m = T/(w_m - \mu)$
over these wealth levels with $w_m=E_m-E_1$ (we apply a 
shift with $E_1$ to ensure that wealth levels are positive 
and have minimal value $w_1=0$). 
We remind that in this approach the thermal evolution of
society wealth in the SSS model is produced by
nonlinear interactions of society agents
according to Eq.~(\ref{eqNLeq1}).
This evolution has two integrals of motion
being the total energy, or total wealth of society
given by $w_{s}= \sum_m w_m \rho_m $,
and the total norm of society population (or number of agents) 
$\eta=\sum_m \rho_m =1$ normalized to unity.
Indeed, the existence of two integrals
of motion is rather natural since
on a scale of one year there is only
a relatively small variation of total wealth $w_s$
and total norm $\eta$.

It is well known that the wealth distribution in the human society, 
e.g. in countries or the whole world,
is characterized by a striking inequality
\cite{piketty1,piketty2,boston} and one may speculate in how far 
it is possible to attribute this inequality to the phenomenon of RJ 
condensation or more generally to the concentration of probability 
on low modes at low temperatures. 

\begin{figure}[t]
\begin{center}
\includegraphics[width=0.46\textwidth]{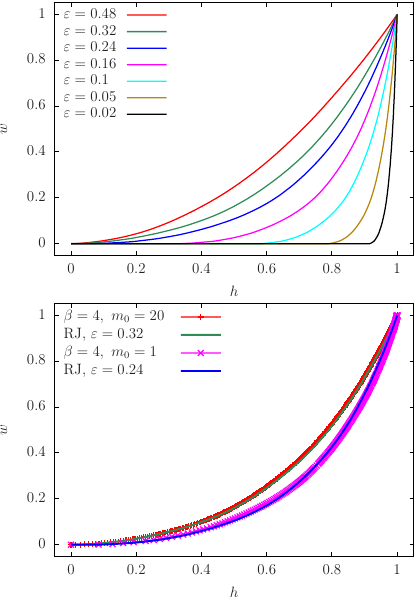}
\end{center}
\caption{\label{fig10}
{\em Top panel:} Lorenz curves using RJ thermalized $\rho_m$ values for the 
spectrum of the SSS model at $W= 8$, $f=0.1$, $\kappa=0,5$, $N=379$
for different rescaled total energies $\varepsilon=w_s/B$. 
The $x$-axis corresponds to the 
cumulated fraction of households ($h$) and the $y$-axis to
the cumulated fraction of wealth ($w$). 
The Gini coefficients $G$ for all curves are
$G=0.96,\,0.9,\,0.81,\,0.71,\,0.55,\,0.46,\,0.33$
(bottom to top).
{\em Bottom panel:} Lorenz curves obtained from numerical 
$\rho_m$ values of the states at $\beta=4$ shown in Fig.~\ref{fig7} at 
maximal time $t=2^{30}$ for the two initial modes $m_0=20$ (red 
$+$ symbols) and $m_0=1$ (pink $\times$ symbols). The green (blue) curve 
corresponds to the Lorenz curve using RJ thermalized $\rho_m$ values for the 
spectrum of the SSS model at $\varepsilon=0.32$ ($\varepsilon=0.24$). 
The $\varepsilon$ value of the green curve corresponds 
$\varepsilon=(E-E_1)/(E_N-E_1)$ where 
$E=\sum_m E_m\rho_m=-1.55$ is the average linear energy 
computed from the numerical $\rho_m$ values of the state at $m_0=20$ 
and $t=2^{30}$. 
The $\varepsilon$ value of the blue curve has been obtained by matching the 
Gini coefficient between numerical data (at $m_0=1$) and the RJ curve. 
Both green and blue curves are identical to the curves of same color in the 
top panel with same $\varepsilon$ values.
Note that the numerical value $\varepsilon_{\rm num.}=0.16$ obtained from 
$E=-2.83$ for the state at $m_0=1$ is rather different 
to $\varepsilon=0.24$ used for the blue curve and corresponds 
to the pink curve in top panel. The Gini coefficients for the 
numerical data are the same as in top panel for the green and blue 
curves. 
}
\end{figure}

A wealth distribution is usually described by the Lorenz curve
\cite{lorenz,boston} that gives the dependence of cumulated 
normalized wealth $w(h)$ on the cumulated normalized fraction of
population or households $h$. 
Here, the hypothetical case of perfect equipartition of wealth corresponds to
the diagonal $w(h)=h$ and the doubled area between diagonal
and the Lorenz curve $w(h)$ determines the Gini coefficient
$0 \leq G \leq 1$  \cite{gini,boston} that describes the overall 
degree of inequality. 
For the world countries the values of $G$ can be found in \cite{wikigini}
being in the range $0.59 < G < 0.90$ in 2021; for the whole world $G = 0.889$.

Let us assume that we have data for population fractions $\rho_m$ with 
wealth levels $w_m$ ($m=1,\ldots, N$, $0=w_1<w_2<\ldots <w_N$) where 
$\rho_m$ may be obtained from the RJ distribution (of some ``wealth'' spectrum 
and given total energy/wealth) or from some other source (empirical 
data, numerically computed values of $\rho_m$ for the models discussed 
in Sections II and III). Then we can construct the Lorenz curve 
as the set of points $(h(m),w(m))$ for $m=0,1,\ldots N$ with 
the partial sums 
$w(m) =  \sum^m_{k=1} w_k \rho_k/w_s$ and 
$h(m) = \sum^m_{k=1} \rho_k$ such that $0 \leq h,w \leq 1$. 
Here the maximal value of $h$ and $w$ is $w(N)=h(N)=1$ 
since $w_s=\sum_m w_m\rho_m$ and $\sum_m \rho_m=1$.

In \cite{wth} a Wealth Thermalization Hypothesis (WTH) was introduced 
according to which the wealth shared in a country or the whole world
is described by the Rayleigh-Jeans thermal distribution (\ref{eqrj})
using certain model spectra of $w_m$. 
This distribution depends on a dimensionless parameter
$\varepsilon = w_s/B$ being the ratio
of the total system wealth $w_s$ on the its dispersion range $B=w_N$ 
determined by its highest value (or $\varepsilon=(E-E_1)/(E_N-E_1)$ if 
$E_1\neq 0$ and with $w_s=E-E_1$). 
At relatively small values of $\varepsilon$  there is the formation of 
an RJ condensate which 
leads to a huge fraction of poor households and a small oligarchic 
fraction which monopolizes a dominant fraction of total wealth thus generating
a strong inequality in human society. It was shown that the WTH concept 
gives a good description of Lorenz curves 
for the wealth distributions in world countries and the whole world,
stock exchange markets of New York, London, Hong Kong, Gross Domestic Product (GDP) of countries
\cite{wth,gdp}.

\begin{figure}[h]
\begin{center}
\includegraphics[width=0.46\textwidth]{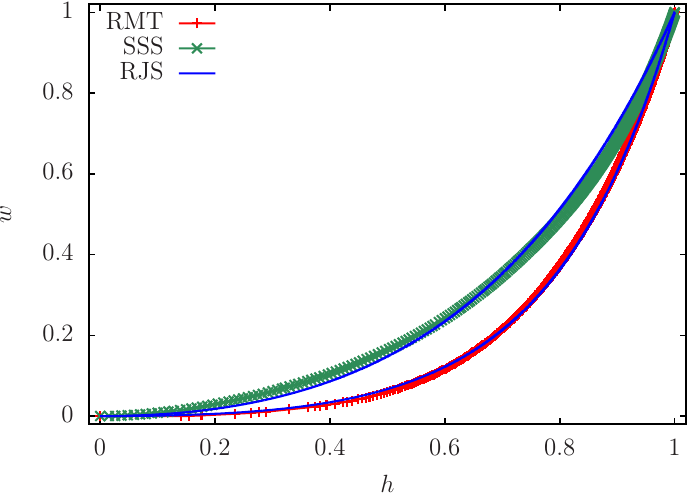}
\end{center}
\caption{\label{fig11}
Lorenz curves for dynamical turbulence for the two 
RMT and SSS models using the data of Fig.~\ref{fig9}. 
The wealth and household variable $w$ and $h$ are computed as explained 
in the text using a shifted reduced spectrum without pumping/absorption modes 
and renormalized of remaining numerical $\rho_m$ values taken from the 
data shown in Fig.~\ref{fig9}. 
The full blue lines correspond to the Lorenz curves of the RJS model 
(using thermalized $\rho_m$ values from a uniform spectrum of $N=10000$ 
energies $E_m=m/N$) with the 
effective rescaled energy $\varepsilon$ being determined to match the Gini 
coefficient $G$ of the data (from RMT or SSS models): 
$\varepsilon=0.202$, $G=0.622$ (RMT) and 
$\varepsilon=0.318$, $G=0.48$ (SSS); $\gamma = \sigma =0.01$ for both models.
}
\end{figure}

In this work in the top panel of Fig.~\ref{fig10}, we present 
several Lorenz curves for the RJ thermal distribution (\ref{eqrj}) 
with the (shifted) eigenvalue spectrum of the matrix $H$ for the SSS model 
(\ref{eqHdef}). 
Each curve corresponds to a given total energy of the system
represented by the rescaled value $\varepsilon =w_s/B$
with $B\approx 8$. The Gini coefficient changes from
$G=0.33$ at $\varepsilon=0.48$ to $G=0.81$ at $\varepsilon=0.1$
and $G=0.96$ at $\varepsilon=0.02$. At $\varepsilon=0.1$
the fraction of poor households that owns 2\% of total wealth is
67\% while the oligarchic fraction of 10\% richest households
owns 63\% of total wealth. Thus we see that the SSS model
with RJ condensate gives a good
description of wealth inequality in the world countries
described in \cite{piketty1,piketty2}.

\begin{figure}[h]
\begin{center}
\includegraphics[width=0.46\textwidth]{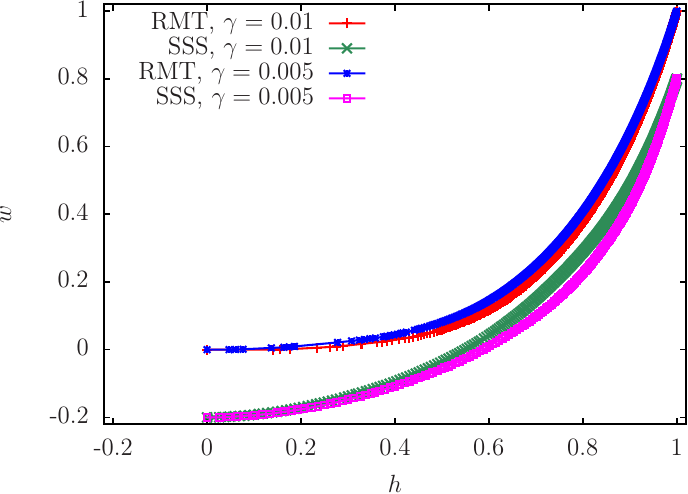}
\end{center}
\caption{\label{fig12}
Same as in Fig.~\ref{fig11},
Lorenz curves are shown with numerical data of $\rho_m$
for the SSS model and the RMT model of KZ from (\cite{ourturb})
for two values of strength pumping $\gamma=0.01$ (as in Fig.~\ref{fig11})
and $\gamma=0.005$ ($\sigma=0.01$ for all curves). 
The curves for SSS are artificially shifted down 
by 0.2 units in $w$ for a better visibility. 
For $\gamma=0.005$ 
we have $G=0.585$, $\varepsilon =0.235$ (RMT);
$G=0.533$, $\varepsilon=0.227$ (SSS); for $\gamma=0.01$
the related values are given in Fig.~\ref{fig11}. 
Here the values of $\varepsilon$ are determined from the RJS model by matching 
the Gini coefficients with the numerical data. The resulting 
RJS Lorenz curves are not shown but they match the numerical curves as 
in Fig.~\ref{fig11}. For both models the Lorenz curves for the 
two $\gamma$ values are rather close. 
}
\end{figure}

The bottom panel of Fig.~\ref{fig10} shows two Lorenz curves obtained 
from the numerical data for $\rho_m$ of the states at $\beta=4$, $m_0=1,20$ 
(see also Fig.~\ref{fig7}) using the (shifted) spectrum of the SSS model for 
the wealth levels. For comparison, we also present two Lorenz curves using 
RJ thermalized $\rho_m$ values (as in top panel) for $\varepsilon=0.32$ and 
$\varepsilon=0.24$. The first value $\varepsilon=0.32$ (with 
green Lorenz curve in both panels) corresponds 
indeed to $E=-1.55$ obtained from the numerical values $\rho_m$ values 
by $E=\sum_m E_m\rho_m$ for the state with initial mode $m_0=20$ and 
it matches very well the direct numerical Lorenz curve (red data points 
in bottom panel of Fig.~\ref{fig10}). This is not astonishing since 
for this state the numerical $\rho_m$ values match indeed quite well the 
thermalized theoretical $\rho_m$ values (see blue line and data points 
in bottom panel of Fig.~\ref{fig7}). 

We mention that for this state ($m_0=20$, corresponding to $\varepsilon=0.32$) 
we have a Gini coefficient $G=0.46$ and 
the fraction of poor households that owns 2\% of total wealth is
18\% while the oligarchic fraction of 10\% richest households
owns 27\% of total wealth. This corresponds to a reduced inequality 
in comparison to the state at $\varepsilon=0.1$ and real country/world 
data of \cite{piketty1,piketty2}.

The other value $\varepsilon=0.24$ (with blue Lorenz curve in both panels) 
has been obtained by matching the Gini coefficient of the numerical 
data for the state with $m_0=1$ (pink data points) with the RJ curve using 
the SSS model spectrum. 
However, here the linear energy $E=\sum_m E_m\rho_m=-2.83$ obtained from 
this state corresponds to $\varepsilon_{\rm num.}=0.16$ and its 
RJ Lorenz curve (only shown in top panel as pink curve) is significantly 
below the numerical Lorenz curve (for $m_0=1$). In top panel of 
Fig.~\ref{fig7}, we indeed see that the RJ distribution of 
$\rho_m$ (blue curve there) 
is somewhat shifted from the numerical data. The reason is that 
for this state many modes $m$ with $E_m>1.3$ have still $\rho_m$ values 
clearly below the thermalized value and they give a significant reduction 
in the computation of $E=\sum_m E_m\rho_m$. However, these large energy 
modes, not yet thermalized, influence the numerical Lorenz curve 
only very slightly in the regime of very rich households (with $h\approx 1$). 
Therefore one can argue that for this state one can reduce the effective 
bandwidth  $B=8\to (2/3)B=5.33$ (of ``well thermalized modes'') 
corresponding to a maximal cutoff energy 
$E_{\rm cut}\approx 1.33$ in the top panel of Fig.~\ref{fig7}. This 
reduction of $B$ gives a rescaling $\varepsilon=0.16\to 0.16/(2/3)=0.24$ 
which is indeed the value for the blue curve obtained by matching 
the Gini coefficient. 

For this state ($m_0=1$, corresponding to $\varepsilon=0.24$) 
we have a Gini coefficient $G=0.55$ and 
the fraction of poor households that owns 2\% of total wealth is
25\% while the oligarchic fraction of 10\% richest households
owns 36\% of total wealth. This corresponds to an ``intermediate'' 
inequality between the case $m_0=20$ ($\varepsilon=0.32$) and 
$\varepsilon=0.1$ and real country/world data of \cite{piketty1,piketty2}.

We also compute Lorenz curves for the case of KZ turbulence 
(see Section III and \cite{ourturb}). Here, we first remove 
first the 4 pumping modes ($m=1,2,3,4$) and the absorption modes 
($m=N-3,N-2,N-1,N$), we apply a shift using the new minimal value $E_5$ 
by $w_m=E_{m+4}-E_5$ for $1\le m\le N-8$ and then we normalize the 
remaining numerical $\rho_m$ values (for $5\le m\le N-4$ 
with subsequent index shift $m+4\to m$ for $1\le m\le N-8$) 
to unity. Note that for the KZT case 
$\sum_m \rho_m\neq 1$ even after removing of pumping/absorption modes. 
Then we use the new shifted reduced spectrum and the 
renormalized $\rho_m$ values 
to compute Lorenz curves as described above (including a recomputation 
of $w_s=\sum_m w_m \rho_m$). 

We applied this procedure to the data of Fig.~\ref{fig9} for the RMT case 
of \cite{ourturb} ($\beta=1$, $N=512$) and the SSS model ($\beta=4$, $N=379$) 
for the steady state distribution of $\rho_m$ in each case. 
The resulting Lorenz curves are shown Fig.~\ref{fig11}. Here 
the fraction of poor households that owns 2\% of total wealth is 
for the RMT (SSS) model 
35\% (16\%) while the oligarchic fraction of 10\% richest households
owns 38\% (32\%) of total wealth.

We also compare 
the numerical Lorenz curves to the Rayleigh-Jeans standard (RJS) model 
(blue curves) introduced in \cite{wth} 
with values of $\varepsilon$ determined by matching the Gini coefficients $G$ 
to the numerical curves (obtained values of  $\varepsilon$ and $G$ 
for both cases are given in the caption of Fig.~\ref{fig11}) and 
the two blue curves for the RJS model match very nicely the numerical curves 
for the RMT and SSS models. 

We remind that the RJS model corresponds to thermalized $\rho_m$ values 
using a simple uniform spectrum 
$E_m=w_m=m/N$ which actually produces Lorenz curves very similar to the 
RJ Lorenz curves of the SSS model (shown in top panel of Fig.~\ref{fig10}) 
since both have a (roughly) constant density of states for a shifted 
spectrum (i.e. with minimal wealth value $w_1\approx 0$) and the different 
values of the bandwidth $B$ do not affect the Lorenz curves (which are 
invariant to global energy rescaling). 
However, for Lorenz curves at small $\varepsilon$ values 
(small $T$ and large $G$) 
the effect of the first few levels is more strongly visible 
(case of RJ condensation) and therefore the precise structure 
of spacings between these first levels may generate some visible differences 
in the Lorenz curves between RJS and RJ for the SSS model. 

We note that for KZ turbulence models the Lorenz curves are 
not very sensitive to the pumping rate $\gamma$ 
as it is illustrated in Fig.~\ref{fig12} which compares the numerical 
curves of Fig.~\ref{fig11} obtained for $\gamma=0.01$ to the case 
of $\gamma=0.005$.
However, the Lorenz curves are sensitive
to the actual spectrum $w_m=E_m-E_1$ of shifted eigenenergies
(or wealth levels of a society) as the difference between the 
RMT and SSS models show. 

We point out several features due to which we consider that the SSS models
described in this work capture  important elements of real
social stratification
and wealth inequality in a society. First, these models
have links between society agents (network nodes)
as they are in real social networks studied in
\cite{newmannets,newman2001,newman2006,newman2006ref84}. 
Second, we introduce the diagonal matrix term
which describes a linear wealth growth
at different society levels
corresponding to the mathematical model
of social stratification in society
that is broadly discussed in social sciences
(see e.g. \cite{marx,lenski,sanders,kerbo}).
Finally and third, we show that
nonlinear interactions between society agents,
of a strength being above a chaos border,
leads to RJ thermalization and condensation
due to dynamical chaos.
The thermalization takes place for various types of nonlinearity
as it was shown in \cite{rmtprl}.
The WTH concept proposed in \cite{wth}
shows that RJ thermalization
naturally explains
the wealth inequality in world countries
which is broadly discussed in economy science \cite{piketty1,piketty2}.
Thus the results obtained here show that
this RJ thermalization and condensation
naturally appear in the mathematical SSS models
due to dynamical thermalization
induced by nonlinearity and chaos.
This gives mathematical grounds for the validity
of wealth inequality in these SSS models
supporting the WTH explanation
of the wealth inequality in the world.

\section{Discussion} 
\label{sec6}

In this work, we introduced a mathematical model of social stratification
of society and studied its properties with a nonlinear interactions
between society members (or agents). In this SSS model
the linear couplings between agents
are described by an adjacency matrix of links
represented by a real network of scientific
collaboration taken from \cite{newman2006}.
The social stratification, known to be present
in a society \cite{marx,lenski,sanders,kerbo},
is described by an additional diagonal
term of the linear Hamiltonian with eigenenergies $E_m$.
These (shifted) $E_m$ energies are interpreted as certain wealth
levels $w_m$ of society with $w_m=E_m-E_1$
counted from the ground state energy $E_1$.
The SSS model includes also 
nonlinear interactions of agents.
The time evolution of the system
is described by a nonlinear Hamiltonian 
with two integrals of motion
being total system energy and total norm.
Our studies show that
above a certain chaos threshold of nonlinearity strength
the nonlinear chaos leads to
dynamical the RJ thermal distribution
over linear eigenmodes with
temperature and chemical potential
determined by the values of two integrals.
At low total energy this RJ distribution has a condensate phase
that  captures a significant fraction of total norm
at low energy modes. 
This RJ condensation has been well studied
theoretically and experimentally
with mulimode optical fibers
\cite{ourfiber,picozziphrep,wabnitz,picozzi1,picozzi2,babin,chris,picozzi3}.
We also show that a model modification for 
energy and norm pumping at low energies and
absorption at high energies in
the SSS model demonstrates features
of the  KZ wave turbulence \cite{zakharovbook,nazarenkobook,galtier}
with an algebraic probability distribution
over system modes.

On the basis of the analogy between
system mode energies and society wealth levels
we show that the Lorenz curves
obtained in the framework of the SSS model
show close similarities with 
the wealth distribution in society \cite{piketty1,piketty2,wth,gdp}.
Thus we consider that the proposed mathematical SSS model
provides a realistic thermodynamic description
of wealth inequality in human societies.

\begin{acknowledgments}
This research has been partially supported through the grant
NANOX $N^\circ$ ANR-17-EURE-0009 (project MTDINA) in the frame 
of the {\it Programme des Investissements d'Avenir, France}. 
This work was granted access to the HPC resources of
CALMIP (Toulouse) under the allocation 2026-P0110. 
\end{acknowledgments}

{\bf The data that supports the findings of this study are available within the article.}


\end{document}